# User Characteristics of Olympic Gold Medallists on Instagram: A Quantitative Analysis of Rio2016

Amirhosein Bodaghi

Federal University of Rio de Janeiro, Department of Computing Science, Centre of Mathematical and Natural Sciences – CCMN, Rio de Janeiro, Brazil
bodaghi@ppgi.ufrj.br

**ABSTRACT**

*The purpose of this study is to examine Olympic champions' characteristics on Instagram to first understand whether differences exist between male and female athletes and then to find possible correlations between these characteristics. We utilized a content analytic method to analyse Olympic gold medallists' photographs on Instagram. By this way we fetched data from Instagram pages of all those Rio2016 Olympic gold medallists who had their account publically available. The analysis of data revealed the existence of a positive monotonic relationship between the ratio of following/follower and the ratio of engagement/follower for men gold medallists, and a strong negative monotonic relationship between age and ratio of self-presenting post of both men and women gold medallists which even take a linear form for men. These findings aligned with the relative theories and literature may come together to help the athletes to manage and expand their personal brand in social media.*

**KEYWORDS**

*Instagram, self-presenting, user characteristics, Olympics, gold medallists*

## 1. INTRODUCTION

In contrast with traditional mainstream media, social media provides athletes with more capabilities to express themselves in a way they prefer most. By sharing of content on social media, athletes interact with their audiences which itself leads to active engagement from the individuals (Frederick, Lim, Clavio, and Walsh, 2012). Rui and Stefanone (2013) found that users who strategically manage their online self-presentation are those who care most about public evaluations and many as well as athletes may fall into this category. Karg and Lock (2014) came to this conclusion that while social media may not directly yield revenue, but by establishing fan communities they confer the advantage of brand awareness to sport entities which itself provides other outcomes. Lebel and Danylchuk (2012) stated that social media has conferred lots of opportunities to athletes to express themselves via social media, so they must be aware of how they present themselves in cyber space. Also, DeAndrea and Walther (2011) analysed Facebook users' self-presentations and concluded each post a person shares has the potential to promote or ruin the cyber character of that person. Especially, it is of great importance for well-known athletes who do not receive much of mainstream media coverage to take advantage of social media to generate publicity, as this is often their only way to make their personal brand (Eagleman, 2013; Parmentier and Fischer, 2012).

As social media evolves into more visual content, interactive engagement and discussion can be driven by sharing of photos and videos (Marshall, 2010). The social media brings freedom for individuals to choose content and come up with different identities (Bullingham and Vasconcelos, 2013). Pegoraro and Jinnah (2012) and Hambrick and Kang (2014) confirm that athletes who properly use social media as a relationship marketing tool to build their online brand would gain brand loyalty. The rapid uptake of Instagram as a marketing platform by the world's most popular brands combined with the potential marketing and branding implications it can have for organizations and individuals merit further exploration of this visual social media platform in the sport context.

Therefore, the purpose of this study is to examine Olympic champions' characteristics on Instagram to first understand whether differences exist between male and female athletes and then to find possible correlations between these characteristics in order to develop an understanding of the ways in which athletes use this medium as a tool to present themselves in cyber space. The results will be explained by related theories and compared with previous findings and the implications may be utilized by athletes to develop their personal branding on online social networks. Furthermore, as we can see from the literature (Casaló et al., 2017; Lee and An, 2018) the antecedents of users represent a lot about their behaviours, and findings of our research about user characteristics can pave the way to dig more in this line of research that incorporates cyber behaviours and has a variety of applications as well in modelling of rumours and information spreading in online social networks (Liu et al., 2019; Bodaghi and Goliaei, 2018; Bodaghi et al., 2019; Bodaghi and Oliveira, 2020).

## 1.1 The structure

In continue, after presenting the literature we begin with the process of data gathering from the Instagram accounts of Olympic gold medallists then we perform a gender breakdown of the Olympic gold medallists' user characteristics such as following/follower, engagement/follower, age and ratio of self-presenting posts. The results will be shown by 2D histograms which clearly reveal the differences between men and women in terms of frequency of user characteristics' values.

## 2. LITERATURE

In a general view, one can look at the topic as a social approach toward mining of Instagram data. Hence, at first, we come up with surveying previous research about social aspects of Instagram then we try to present recent research findings, which are focused on user characteristics to analyse social behaviours on Instagram.

### 2.1 Social Aspects of Instagram

Having regard to the Sociotechnical aspects of using social networks, a series of research has been particularly done on Instagram. Souza et al. (2015) study the collective behaviour of sharing selfies on Instagram and present how people appear in selfies and which patterns emerge from such interactions. Pittman and Reich (2015) suggest that Image-Based social media such as Instagram could reduce the loneliness of users. Hochman and Schwartz (2013) use cultural analytics visualization techniques to study Instagram Images, with the aim of tracing cultural visual rhythms. Silva et al. (2013) characterize users' behaviour in the system showing that there are several advantages for large-scale sensing by photo sharing on Instagram. Ferwerda et al. (2015) study to infer personality traits from the way users take pictures and apply filters to them. Araujo and Correa (2014) investigate the user practices on Instagram, in terms of date of posting and some social behavioural attitudes. Lup et al. (2015) explore associations among Instagram use, depressive symptoms and strangers followed.

By exploring the literature of research on social aspects of Instagram, we notice that the analyses of past research are particularly based on the posts' photos and rarely use other information to discover a high-level concept in human behaviour, however in this research we lay our analysis on both photo-driven and text-driven information to explore the existence of any trace to human behaviours.

### 2.2 User Characteristics

Jang et al. (2015) analyse like activities in Instagram. They provide an analysis of the structural, influential and contextual aspects of like activities from the test datasets of users on Instagram. Ferrara et al. (2015) focus on online popularity and topical interests through Instagram. Their analysis provides clues to understand the mechanisms of users interacting in online environments and how collective trends emerge from individuals topical interests. Bakhshi et al. (2014) investigate the influence of faces in Instagram photos in terms of the number of likes and comments those photos attract. As we see a variety of user characteristics including the number of likes and comments for each user's post in average, age, gender and so on have been used for different research items. However, here we use the ratio of some user characteristics such as like/follower and follower/following, each one as a separate variable to investigate its possible relationships with the frequency of self-presenting posts.

### 2.3 Age and gender

Some researchers focus on user characteristics to analyse social behaviours on Instagram, because they form a window through which users' cyber behaviours could be tracked. Sheldon and Bryant (2016) investigate motives for using Instagram and its relationship to contextual age and narcissism. Jang et al. (2015) show the possibility of detecting age information in user profiles by using a combination of textual and facial recognition methods. Tifferet and Vilnai-Yavetz (2014), examined gender differences in Facebook self-presentation by evaluating components of profile and cover photos. They found profile photos on Facebook differ according to gender. Males' photos came with using objects or formal clothing and risk taking, while females' photos dealt with familial relations and emotional expression. Pamara et al. (2015) apply socioemotional selectivity theory to explain differences in the size and composition of Facebook users. Their findings suggest increasing selectivity of Facebook social partners with age. Compared to younger adults, friend networks of older adults are smaller but contain a greater proportion of individuals who are considered to be actual friends.

### 2.4 Self-Presentation of Athletes

A little research has been focused on athlete self-presentation on social media platforms, and results revealed that athletes engage in backstage performances to discuss their private lives and engage with individuals (Burch et al., 2014, Hambrick et al., 2010, Lebel and Danylchuk, 2012). Moreover, Burch et al. (2014) and Lebel and Danylchuk (2012) studied gender differences across self-presentation and found no differences in self-presentation across gender. Up to this date the major of research on athletes' self-presentations via social media has focused on Twitter and Facebook. Geurin-Eagleman and Clavio (2015) investigated the Facebook accounts of specialty sport athletes, characterized as those who were not covered in the media and must rely on their own endeavour to be seen, and mainstream sport athletes, characterized as those who received mainstream media coverage. Geurin and Burch (2016) presented a gender-based analysis of Olympic athletes' self-presentation on Instagram in order to develop an understanding of the ways in which athletes use this medium as a communication and marketing tool to build their personal brand.

Despite all the researches have been centred around the presence of athletes on social media, up to this date no sound research has been conducted to investigate possible relationships between athletes' characteristics on Instagram particularly with themes of gender and age. However, our findings which are based on a measurement study prove the existence of such relationships that would pave the way to advance further investigations about the athletes' cyber characters.

## 3. METHOD

We utilized a content analytic method to analyse Olympic gold medallists' photographs on Instagram. As indicated by Riffe, et al. (2005), content analysis is an efficient and replicable method for examining content, both written and visual. The Olympic of Rio 2016 was held for 16 days from 5 to 21 August. There were 42 summer sports with total of 306 events. The number of gold medals were given to the individual champions (excluding team sports) were 226, and since some athletes were champion in more than one event, the total number of individual gold medallists is 213, 95 woman champions and 118 man champions (https://www.olympic.org), from which 151 (85 men and 64 women) gold medallists had a publicly available Instagram account (Table 1).

Table 1: Individual Events and Gold Medallists of Olympic Rio2016

| Sports With Individual Events | Men Individual Events | Women Individual Events | Gold Winner (Men) | Gold Winner (Women) | Gold Winners with active Instagram Account | |
|---|---|---|---|---|---|---|
| | | | | | Men | Women |
| Archery | 1 | 1 | 1 | 1 | 0 | 0 |
| Athletics | 22 | 21 | 20 | 20 | 16 | 15 |
| Badminton | 1 | 1 | 1 | 1 | 0 | 1 |
| Boxing | 10 | 3 | 10 | 3 | 8 | 3 |
| Canoeing | 6 | 3 | 6 | 3 | 4 | 1 |
| Cycling | 7 | 7 | 6 | 7 | 5 | 7 |
| Diving | 2 | 2 | 2 | 2 | 0 | 0 |
| Equestrian | 2 | 1 | 2 | 1 | 0 | 1 |
| Fencing | 3 | 3 | 3 | 3 | 2 | 2 |
| Golf | 1 | 1 | 1 | 1 | 1 | 0 |
| Gymnastics | 8 | 7 | 7 | 5 | 4 | 5 |
| Judo | 7 | 7 | 7 | 7 | 6 | 4 |
| Pentathlon | 1 | 1 | 1 | 1 | 1 | 1 |
| Rowing | 1 | 1 | 1 | 1 | 1 | 0 |
| Sailing | 3 | 2 | 3 | 2 | 3 | 2 |
| Shooting | 9 | 6 | 8 | 6 | 2 | 2 |
| Swimming | 14 | 14 | 12 | 11 | 12 | 8 |
| Table Tennis | 1 | 1 | 1 | 1 | 0 | 1 |
| Taekwondo | 4 | 4 | 4 | 4 | 2 | 2 |
| Tennis | 1 | 1 | 1 | 1 | 1 | 1 |
| Triathlon | 1 | 1 | 1 | 1 | 1 | 1 |
| Weightlifting | 8 | 7 | 8 | 7 | 6 | 3 |
| Wrestling | 12 | 6 | 12 | 6 | 10 | 4 |
| Sum | 125 | 101 | 118 | 95 | 85 | 64 |

We examined just individual gold medallists in this study because athletes who won gold medals in team sports in the Olympics often compete for other professional teams or leagues in non-Olympic years. Thus, those athletes take advantage of these teams' endeavors and the attention received by the teams/leagues. While individual gold medallists often rely on their own efforts to establish their personal brand and to remain relevant in the intervals between Olympic Games. Then for each user, we gathered the following information: 1-the number of posts; 2-the number of followers; 3-the number of followings; 4-the max number of likes and comments among the last ten posts; 5-the number of self-presenting posts (posts in which the champion face is detectable) from the previous ten posts; 6- the number of pure self-presenting posts (posts in which the champion stands alone) from the previous ten posts; 7- gender; 8- age.

We have only taken photo posts into account and if there were any videos or multiple-pictures (as one post) in the last ten posts then we went back to more previous posts till we found ten photo posts. Another point to mention is the notion of "selfie" which despite of its popularity in Instagram, it is the least popular photograph athletes include on their feed (Smith and Sanderson 2015). Thus, to overcome the limiting frame of selfie which solely focusing on the subject of the photograph, we took into account all photos in which the athletes recognizable as self-presenting post. Indeed, to count the number of self-presenting posts among the last ten posts which is a kind of media-driven information, we counted all posts in which the user's face is recognizable, while for counting pure self-presenting posts, the condition of being solitary in the photo was added.

### 3.1 Data and Codes

The process of gathering information by the introduced method took four days from 9 to 12 August 2019 and all done by human referees with meticulous examining and exploiting data from each Instagram account. The dataset is publicly available on the internet. Also, the codes for data analysis which we will get to in the next section, all have been written by python and are publicly available on the internet.

## 4. RESULTS

### 4.1 Gender Breakdown of the characteristics

In this section we aim to find any difference between characteristics of men and women gold medallists. To have the measured characteristics more comparable, we use them in the form of ratio to each other rather applying them merely by their own. Thus, the characteristics we use in this analysis would be as follows: 1- ratio of following/follower, 2- ratio of engagement (like+comment)/follower, 3- ratio of self-presenting posts among the last ten photo posts, 4- ratio of pure self-presenting posts among the last ten photo posts, 5- ratio of pure self-presenting posts/ self-presenting posts, 6- age, 7- gender.

To have better comparisons of the data in an obvious compact form, we apply 2 dimensional histograms. Figure 1 shows the results for the ratio of following/follower and age in both men and women gold medallists. As it is shown, the value of following/follower less than 0.2 is dominant for both men and women and this fact is of no surprise since champions are the centre of attention and their accounts achieve lots of followers without any obligation to follow back the followers. Moreover, we see that champions of 24 to 33 years of age have almost swept the gold medals for both men and women events. However, it should be considered that this research is conducted almost 3 years after the Olympic Rio2016 so those champions took gold medals when they were 3 years younger.

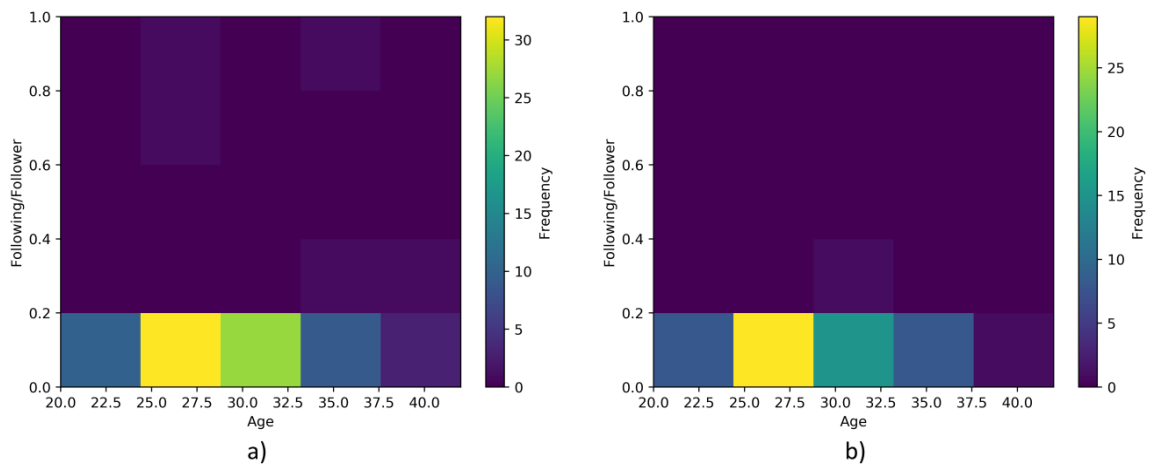

Figure 1. 2-D histogram of Following/Follower and Age. a) for men, b) for women. The squares with warm colours denote areas with more frequency in the dataset. A similar pattern can be seen for men and women, since most of gold medallists have a following/follower ratio less than 0.2 while their ages fall in the range of 24-33.

When we examine the relation of age with the ratio of engagement/follower, it turns out that men champions at the age of 28.5-33 attain the most engagement out of their followers while the same happens for women champions at the age of 24-28.5 (Fig. 2). Furthermore, for both men and women champions, almost less than 0.3 of their followers bother to like or comment their posts, thus to find champions who get a higher level of engagement would not be easy.

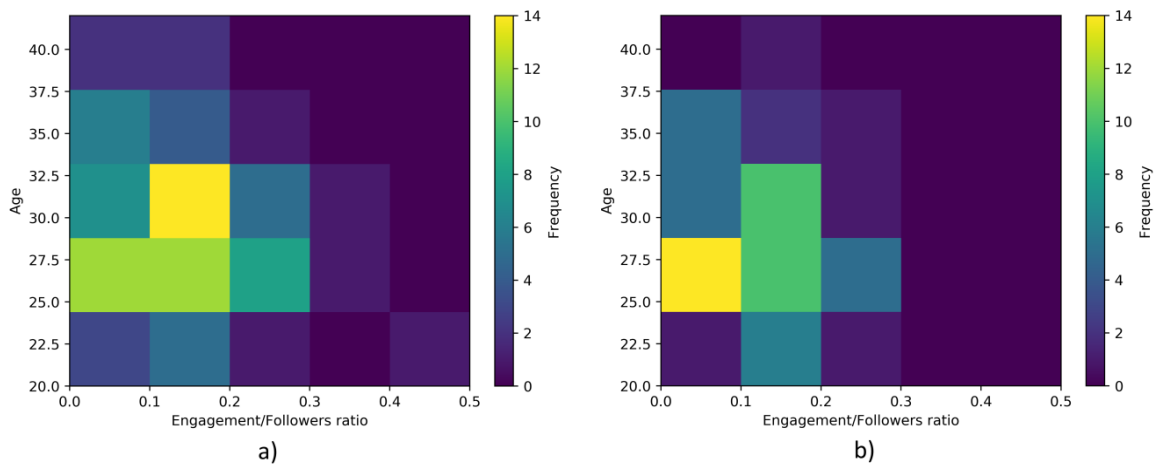

Figure 2. 2-D histogram of Engagement/Followers and Age. a) for men, b) for women. The squares with warm colours denote areas with more frequency in the dataset. Most of the men gold medallists are at the age of 29-33 with engagement/followers ratio between 0.1 and 0.2 while their female counterparts mostly are at the age of 24-29 with engagement/followers ratio between 0 and 0.1.

Having the age characteristic in one hand, we can now compare men and women champions in terms of the ratio of pure self-presenting posts (Fig. 3). By looking at the most crowded range i.e. 24 to 33, it can be seen that men champions tend to share less pure self-presenting posts than women do. This conclusion could also be drawn saliently from the 2D histogram of ratio of engagement/follower and ratio of pure self-presenting posts/ self-presenting posts (Fig. 4). Indeed, when it comes to self-presenting posts of Olympic champions, chances are women be alone in the photo, while men stand with their friends or fans.

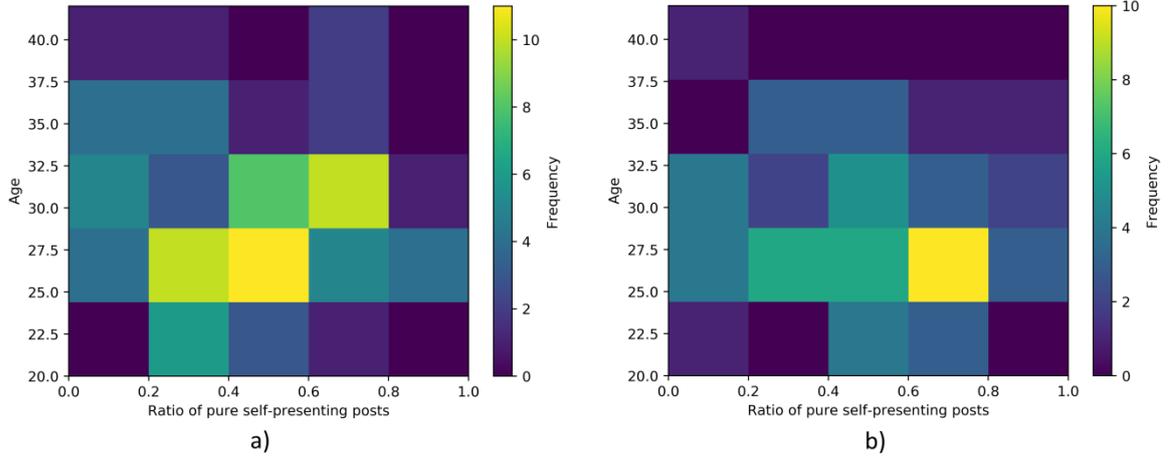

Figure 3. 2-D histogram of the ratio of pure self-presenting posts and Age. a) for men, b) for women. The squares with warm colours denote areas with more frequency in the dataset. The ratio of pure self-presenting posts for men reaches its most frequency at the range of 0.4 to 0.6 while the same for women befalls at the range 0.6 to 0.8.

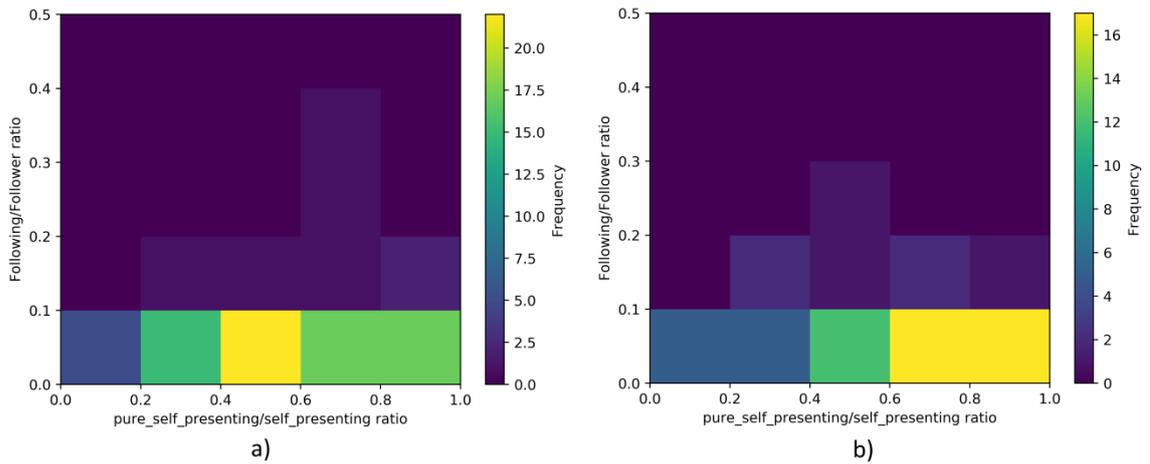

Figure 4. 2-D histogram of the ratio of pure-self-presenting/self-presenting and Fallowing/Follower. a) for men, b) for women. The squares with warm colours denote areas with more frequency in the dataset. The higher ratio of pure-self-presenting/self-presenting posts for women indicates they share more pure self-presenting posts out of their self-presenting posts than men do.

### 4.2 Correlations between the characteristics

In this section we examine the possible relationships between different characteristics of gold medallists. To this aim we computed Pearson and Spearman correlations for each pair of characteristics in three categories 1-general (both men and women), 2-only men, and 3-only women. Table 2 shows the achieved results including coefficient correlations and p-values for each pair of characteristics. To be able to ignore the null hypothesis with high probability, we set $\alpha = 0.05$ in $P-Value < \alpha$, and by doing so, the meaningful relationships get bold in the table.

Table 2: Spearman and Pearson Correlations between the characteristics

| | | | FOLLOWING/FOLLOWER | | ENGAGEMENT/FOLLOWER | | AGE | | RATIO OF SELF-PRESENTING POSTS | | RATIO OF PURE SELF-PRESENTING POSTS | | RATIO OF PURE SELF-PRESENTING POSTS/ SELF-PRESENTING POSTS | |
|---|---|---|---|---|---|---|---|---|---|---|---|---|---|---|
| | | | PEARSON | SPEARMAN | PEARSON | SPEARMAN | PEARSON | SPEARMAN | PEARSON | SPEARMAN | PEARSON | SPEARMAN | PEARSON | SPEARMAN |
| FOLLOWING/FOLLOWER | GENERAL | COEFFICIENT | | | 0.04341 | 0.27341 | 0.10305 | 0.06925 | -0.0931 | -0.1125 | -0.0288 | -0.06235 | 0.00935 | 0.04878 |
| | | P-VALUE | | | 0.59910 | 0.00074 | 0.21105 | 0.40135 | 0.25853 | 0.1717 | 0.72655 | 0.4499 | 0.9098 | 0.5546 |
| | WOMEN | COEFFICIENT | | | -0.00783 | 0.147273 | 0.17983 | 0.15819 | -0.1319 | -0.20706 | -0.16203 | -0.22668 | -0.08201 | -0.05785 |
| | | P-VALUE | | | 0.95143 | 0.24939 | 0.15844 | 0.21561 | 0.30255 | 0.10345 | 0.204512 | 0.074 | 0.52279 | 0.65241 |
| | MEN | COEFFICIENT | | | 0.03727 | 0.3533 | 0.09364 | -0.00465 | -0.117 | -0.0315 | 0.00885 | 0.03032 | 0.0484 | 0.10195 |
| | | P-VALUE | | | 0.73328 | 0.0008 | 0.391 | 0.966 | 0.2833 | 0.77336 | 0.93554 | 0.78168 | 0.658 | 0.35025 |
| ENGAGEMENT/FOLLOWER | GENERAL | COEFFICIENT | | | | | -0.10458 | -0.09583 | 0.05498 | 0.09844 | 0.0589 | 0.05553 | 0.04376 | 0.02264 |
| | | P-VALUE | | | | | 0.20432 | 0.2449 | 0.5054 | 0.2323 | 0.4754 | 0.5011 | 0.5961 | 0.784 |
| | WOMEN | COEFFICIENT | | | | | -0.0168 | -0.05121 | 0.03421 | -0.04806 | -0.07782 | -0.05292 | -0.0801 | -0.01844 |
| | | P-VALUE | | | | | 0.8955 | 0.69014 | 0.79005 | 0.70835 | 0.5443 | 0.6803 | 0.53217 | 0.88592 |
| | MEN | COEFFICIENT | | | | | -0.1704 | -0.1436 | 0.0563 | 0.19045 | 0.14184 | 0.150517 | 0.12639 | 0.1036 |
| | | P-VALUE | | | | | 0.11654 | 0.187 | 0.6065 | 0.079 | 0.1926 | 0.1665 | 0.2461 | 0.342 |
| AGE | GENERAL | COEFFICIENT | | | | | | | -0.3922 | -0.2761 | -0.1562 | -0.11374 | -0.0256 | -0.018 |
| | | P-VALUE | | | | | | | 0.000007 | 0.0006 | 0.05709 | 0.16719 | 0.756 | 0.82748 |
| | WOMEN | COEFFICIENT | | | | | | | -0.35207 | -0.107 | -0.22442 | -0.1948 | -0.1205 | -0.17536 |
| | | P-VALUE | | | | | | | 0.0046 | 0.40386 | 0.077014 | 0.12591 | 0.34656 | 0.16921 |
| | MEN | COEFFICIENT | | | | | | | -0.44986 | -0.4023 | -0.08899 | 0.05089 | 0.0697 | 0.0977 |
| | | P-VALUE | | | | | | | 0.00001 | 0.00001 | 0.41515 | 0.6416 | 0.52309 | 0.37076 |

The acceptable value of Spearman correlation between the characteristics of Following/Follower and Engagement/Follower for general users indicates a considerable positive monotonic relationship between them, in which women do not take any share, since all its weight falls in the men side. Moreover, the lack of any Pearson correlation rejects the premise of being linear for this relationship.

When it comes to the characteristics of age and ratio of self-presenting posts, significant relationships emerge out of both Pearson and Spearman correlations. The negative coefficients of these correlations accompanied by tiny P-values indicate a remarkable inverse linear and monotonic relationship between age and ratio of self-presenting posts. These strong relationships can be seen in both men and women categories too, except for women in Spearman correlation in which P-value remains above 0.05 and ushers a non-monotonic linear correlation between their age and the ratio of self-presenting posts.

Aside for the strong relationships we mentioned above, if we could tolerate a little bit violation from the bound of $\alpha \leq 0.05$ (for instance $\alpha \leq 0.08$) we would find other relationships which may not be as strong as aforementioned but also considerable. The correlations' coefficient of these relationships are underlined in the table 1. First, we see a negative Spearman correlation between the characteristics of Following/Follower and Ratio of pure self-presenting posts which signals an inverse monotonic relationship between them. Second, it can be seen there is a positive Spearman correlation between the characteristics of Following/Follower and Ratio of pure self-presenting posts which indicates a positive monotonic relationship between them. Also, there is a negative Spearman correlation between age and ratio of pure self-presenting posts in the general category which has been inherited from the women gold medallists.

The salient findings of this section can be summed as follows:

- F1: Women gold medallists not only tend to share more self-presenting posts relative to men gold medallists but also their rate of pure self-presenting posts is higher.
- F2: There is a positive monotonic relationship between the ratio of following/follower and the ratio of engagement/follower for men gold medallists.
- F3: For both men and women gold medallists there is a strong negative monotonic relationship between age and the ratio of self-presenting posts which even takes a linear form for men gold medallists.

## 5. DISCUSSION

The results from this study offer several areas of discussion and implications that can be utilized by athletes and their sponsors as well as sport researchers. In the following we will try to discuss each of the main findings in the light of related theories and previous findings of the research context.

## 5.1 Gender differences of self-presenting and theory of evolutionary psychology (F1)

The first finding indicates the higher rate of self-presenting photos for women gold medallists relative to their male counterparts. This finding is aligned closely with the result of a similar research about Olympic athletes conducted by Geurin-Eagleman and Burch (2016). They demonstrated that female athletes appear in their Instagram photos more often than male athletes, while the male athletes are more successful in eliciting engagement from their fan than the female athletes. However, their findings were limited to only 8 selected Olympic athletes. We use evolutionary psychology – a theory which holds many assumptions regarding gender differences – to explain this finding. The core idea in evolutionary theory is natural selection which states traits that support survival and reproduction have a higher chance to be passed to next generations (Gazzaniga, Ivry and Mangun, 2009). On the other hand we have the term "self-concept" which can be defined as a set of attributes, attitudes, and values that individuals believe defines themselves (Berk, 2009). To make sure survival and reproduction are fulfilled, the Mother Nature has given different roles to men and women by which they define themselves.

Women, as the main caregivers, develop skills for attachment, sensitivity and communication. Men, on the other hand, had to display power and dominance in order to maintain their status (Dovidio, Brown, Heltman, Ellyson, and Keating, 1988) Also, a series of research on visual portrayals found that female athletes were featured more in gender appropriate sports and in posed or passive photos rather than action shots, which reinforced stereotypical gender roles (Fink and Kensicki, 2002; Hardin et al., 2005). Thus, along with their defined self-concept, women tend to convey emotions, so they exhibit more facial cues which requires a close shot (mostly pure self-presenting posts) or at least a photo with friends, family or fans (mostly self-presenting posts). While, men tend to present their dominance and possessions which can also be shown by not making themselves as the subject of photo, for instance a runner posts photos of his shoes collection or a weight lifter post a photo of another weight lifter (usually his rival) who has put all his strengths and efforts to lift a weight that is lighter than his record, even though he admires and encourages that rival with his tribute, but in fact he put the message of his dominance in the mind of his followers.

## 5.2 Gender differences of self-presenting and narcissism (F1)

The frequency of self-presenting posts can be interpreted as narcissism which is a personality trait that entails a person having an exaggerated self-concept and a desire to be admired (Buffardi and Campbell, 2008). There is a couple of research which focus on narcissism in social networks (Sheldon and Bryant, 2016; Kapidzic, 2013) and some which claim the existence of a positive relationship between narcissism and the frequency of selfie posts (Sheldon, 2016). Narcissism is positively related to using Instagram to appear cool and for surveillance of others and also significantly related to the number of time participants spend editing the photos before posting them on Instagram (Sheldon, 2016). However, we little know about how these gender differences of self-presentation should correctly map into different levels of narcissism in men and women particularly in the realm of sports. The finding of this research which has been conducted on men and women athletes who are at the same rank (both are Olympic gold medallists) may outline the biased norms of self-presentation on social media with respect to the gender. Even though, now we have this primitive understanding that women athletes who have more self-presenting posts relative to men athletes are not necessarily more narcissist but much more research would be required to clarify the variety of forms by which narcissism finds its way into the cyber character of athletes, and then quantifying their level of narcissism.

### 5.3 Engagement and the ratio of following to follower (F2)

As we have seen in the literature, athletes tend to share their backstage performances, and based on this, one might reasonably posit that posting more personal life photos in which the athlete appeared would encourage greater fan engagement, which is important in building a personal brand, as continued audience engagement in the form of interactivity leads to brand loyalty with consumers (Hambrick and Kang, 2014). Geurin A. and Burch, L. (2016) found that athletes wishing to raise their public profile and build their personal brand should experiment with posting a wider variety of photo types.

Our findings evince another promising way to boost engagement from the athletes' followers and that is to increase the ratio of following/follower. In other words, gold medallists who want to get more out of their followers better to follow others too. However, our finding showed that this strategy only works for men athletes, which itself opens a new subject to discuss why women gold medallists do not benefit from it. Even though answering this question requires further research, but we argue that the answer may have roots in the reasons followers come along with to follow the athletes. Indeed, reciprocal relationships which take the form of follow and follow back on social media, befalls on situations in which both sides see their benefit in mutual friendship almost at the similar way. Women athletes particularly at the high rank of gold medallists, aside from their sport fame, benefit a variety of privileges such as sexuality and attractiveness particularly for their men followers who engage in their posts without any expectations to be followed back, or at least, behaving the same for women followers of men gold medallists may not be prevalent as well.

### 5.4 Aging and self-presentation (F3)

Roberts, Edmonds, and Grijalva (2010) came to this conclusion that narcissism decreases with age, since the narcissistic characteristic of not making commitments to others goes against the normative pathways. Another study on NPI narcissism found a steady decrease in narcissism between age 15 and 54, with a small increase after age 55 (Foster, Misra, and Reidy, 2009). Litchfield and Kavanagh (2018) showed male athletes tend to be presented in an active pose and less likely to be presented in a passive pose compared to female athletes. Krane et al. (2011) examined the desired type of sport images young females preferred to see, to find authenticity and reflection of self were characteristics of positively viewed photos.

Add to this, our findings show that with aging gold medallists lose their interest to share their self-presenting posts, however, this decrease for men takes a linear form while for women only remains monotonic. We argue the reason resides in the difference of gender-basis tendencies for self-presentation on social media. In fact with aging men gold medallists who used to share their self-presentations as active posts with the spirit of an unbeatable man, now feel less strength to be active like the past. However, women gold medallists who were more satisfied with presenting of themselves in passive posts would not run out of interest in self-presentation by aging as much as their male counterparts.

## 6. CONCLUSION

By doing an Instagram analysis on Olympic gold medallists in terms of their user characteristics, we found an early understanding of their cyber behavioural diversity. In fact, the user characteristics open a window for us, through which we can study human behavioural features more precisely. The findings serve as a guide for sport researchers seeking to better understand athletes' use of social media, particularly Instagram, to interact with their followers and build the athlete's personal brand which involves sponsorship and other business/promotional opportunities.

## 6.1 Limitations

In some rare cases, we faced with posts that were not clear to be a pure self-presenting posts, for example in some photos the gold medallists were in the photo without any friend or fan close to them, however there were other people in the photo placed in different locations, which seemed there had been no intention to have them in the photo, thus we considered those photos as pure self-presenting posts. Also, a few posts were the paintings that depicted the gold-medallists' faces as the only subject of the photos, and we considered them as pure self-presenting posts.

One may also claim that the Instagram accounts we analysed were not the real account of the gold medallists but of someone else. Almost, all of the gold medallists accounts we investigated had the blue mark of official registering in Instagram and for those few gold medallists who had not their Instagram account officially registered we studied all their posts to make sure those accounts belong to them and for this we set the basis on the subject of photo posts, because it is highly likely that when one adult was shown in the photo, the photo belonged to that user (Tifferet and Vilnai-Yavetz, 2014).

## 6.2 Future works

Athlete's cyber behaviour could also be tracked with a longitudinal study to determine whether and how athletes' personal branding efforts and strategies change over time. Also, aside from content analysis of athletes' self-presentations, we can interview the athletes themselves to gain further insight into their cyber behaviour. It would also be beneficial to examine fans' impressions of athletes' social media behaviours to develop a better understanding of their interest, and to know the impact of athletes' posts on the audiences. While this study took into account only the number of photo comments with the same weight as we considered for the number of likes, however, a further research can qualitatively analyse these comments and assign different weights to them relative to the likes and by doing so it would be possible to achieve a more precise measure of engagement.

**Authors**

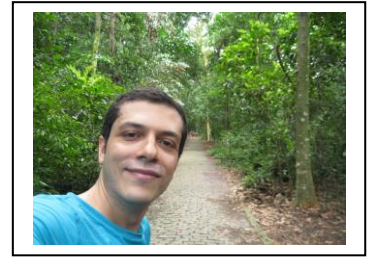

Amirhosein Bodaghi is a postdoc researcher in the Federal University of Rio de Janeiro. The focus of his research is computational social science and social media.